# GAP: Enhancing Semantic Interoperability of Genomic Datasets and Provenance Through Nanopublications

Matheus Pedra P. Feijoó[1][0000-0001-9582-4595], Rodrigo Jardim[3][0000-0002-0943-5356],
Sergio Manuel Serra da Cruz[1,2][0000-0002-0792-8157],
Maria Luiza Machado Campos[1][0000-0002-7930-612X]

[1] Universidade Federal do Rio de Janeiro (PPGI/UFRJ), Rio de Janeiro, Brazil
[2] Federal Rural University of Rio de Janeiro (UFRRJ), Seropédica, Brazil
[3] Oswaldo Cruz Foundation (LBCS/IOC), Rio de Janeiro, Brazil
`feijoo@ufrj.com, mluiza@ppgi.ufrj.br, serra@ppgi.ufrj.br`

**Abstract.** While the publication of datasets in scientific repositories has become broadly recognised, the repositories tend to have increasing semantic-related problems. For instance, they present various data reuse obstacles for machine-actionable processes, especially in biological repositories, hampering the reproducibility of scientific experiments. An example of these shortcomings is the GenBank database. We propose GAP, an innovative data model to enhance the semantic data meaning to address these issues. The model focuses on converging related approaches like data provenance, semantic interoperability, FAIR principles, and nanopublications. Our experiments include a prototype to scrape genomic data and trace them to nanopublications as a proof of concept. For this, (meta)data are stored in a three-level nanopub data model. The first level is related to a target organism, specifying data in terms of biological taxonomy. The second level focuses on the biological strains of the target, the central part of our contribution. The strains express information related to deciphered (meta)data of the genetic variations of the genomic material. The third level stores related scientific papers (meta)data. We expect it will offer higher data storage flexibility and more extensive interoperability with other data sources by incorporating and adopting associated approaches to store genomic data in the proposed model.

## 1 Introduction

Datasets and reproducibility of research play a crucial role in modern data-driven research. Scientific data management has become increasingly complex and is gaining traction in the research community, mainly when spotlighting data and metadata's share, reuse, and interoperation, particularly for machine-actionable processes [1]. Genomic databases are classic examples of this scenario.

Researchers often upgrade their databases with diverse data and metadata to map as sequence new genes, fomenting new biological investigations [2]. These databases store either partial sequences of genes or complete genomes of organisms. However, data reusability-related issues challenge researchers. Sometimes data are unnecessarily duplicated, inconsistent, inaccurate, incomplete, and even obsolete, to name a few of these issues [3, 4]. Consequently, these drawbacks aggravate whenever researchers need to find, access, interoperate, and reuse data using machine-actionable processes, requiring a fully semantic-aware scenario [5].

Our work investigates two core issues related to data reuse. First, machine-driven processes cannot understand the real meaning of stored data in repositories, generating an inflexible interoperable scenario. Second, the lack of data standardisation and provenance [6, 7] makes understanding the data more error-prone and time-consuming. This paper claims that we can address both issues with an innovative approach to managing data provenance while ensuring semantic interoperability based on nanopublication technologies; we present the Genome Assembly nanoPublication (GAP) approach.

Nanopublications [8] (a.k.a nanopub) are a formalized and machine-readable way of communicating the smallest possible units of publishable information. A nanopub is the materialisation model of Linked Data (LD) concepts representing small statements attached with data provenance and metadata. According to [9], it creates a uniform, self-supporting, and machine-readable information ecosystem. Nanopubs is a novel and reliable approach for extending scientific insights, notably in the biomedical fields [10]. However, the current nanopubs suffer from the absence of the data provenance of the assertions and the nanopub itself, like incorrect authorship information [11].

This article discusses the feasibility of a novel data model to enhance the control of genomic data by injecting descriptive and structural semantically enriched metadata. The devised model uses convergent approaches and techniques such as controlled vocabularies, ontologies, W3C PROV standard [12], and FAIR data principles [5] to mitigate the nanopubs related issues. As proof of concept, we have created an operational prototype of GAP to transform genomic data into nanopubs. In order to evaluate the proposal, the nanopubs were confronted with the real-world scenario of reusing genomic (meta)data and other well-known nanopubs repositories to verify the potential of data provenance and interoperability enrichment in this machine-actionable context.

Our computational experiments consider the GenBank Assembly Database (GBAD) as a use case to compose this novel data model. Among all GenBank databases, GBAD is one of the most used [13]. The data it stores are manually published and curated by researchers or scientific organisations. These publishers usually store data related to the assembly of organisms, containing the composition of assembled genomes, additional metadata, statistical reports, and genomic sequence data [14].

The paper is organised as follows. First, we present the background and discuss related works in Section 2. Then, in Section 3, we detail the proposed approach, the Genome Assembly nanoPublication (GAP) data model. Next, we report our implementation and discussion in Section 4 and conclude the paper in Section 5.

## 2  Background and Related Work

### 2.1  FAIR Data Principles and Nanopublications

The FAIR data principles (FDP) [5] focus on reusing data and metadata of any kind. These principles stress the role of Open Science with a focus on '(meta)data', where authors use this term in cases that apply to metadata and data. FAIR stands for: Findable, Accessible, Interoperable, and Reusable. The principles based on these characteristics, up-to-date data resources, tools, vocabularies, and foundations should manifest to support exploration and reuse by third parties through the Web [5].

Reusability and interoperability are quintessential goals for Web data: data resources should be explicitly designed to be reused by either humans or machines. The interoperability in data infrastructures maximises the value of information artefacts retrieved from different data silos. However, a much broader and deeper analysis must have a reliable and solid data association.

Several proposals report the lack of semantic interoperability, undocumented data models, poor provenance, and data reuse. In the context of scientific repositories, the nanopublication model can address some of these problems.

Nanopubs follow the FDP and exploit the LD to represent any digital object using minimal statements with its context and provenance. A typical nanopub uses the Resource Description Framework (RDF), persistent identifiers, use licenses, the Web Ontology Language (OWL), and the nanopub schema. Initially, a nanopub consists of four RDF graphs: head, assertion, provenance, and publication info [8].

Figure 1 represents a nanopublication with the following assertion of "Malaria is transmitted by mosquito." The assertion is the core section in a nanopub, represented by at least one RDF triple. This assertion constitutes two concepts (Malaria and Mosquito) and a relationship (Transmitted by). Therefore, the RDF statement triple is displayed as Subject (Malaria), Predicate (Transmitted by), and Object (Mosquito).

```
@prefix cw: <http://conceptwiki.org/index.php/Concept> .
@prefix orcid: <http://orcid.org> .
@prefix np: <http://www.nanopub.org/nschema#> .
@prefix pav: <http://purl.org/pav/> .
sub:Head{
    this: np:hasAssertion sub:Assertion ;
        np:hasProvenance sub:Provenance ;
        np:hasPublicationInfo sub:Pubinfo ;
        a np:Nanopublication .
}
sub:assertion{
    cw:malaria cw:Trasmitted_by cw:mosquitoes.
}
sub:provenance{
    sub:assertion pav:authoredBy cw:BobSmith ;
        pav:createdOn "2008-08-05"^^xsd:date .
}
sub:pubinfo{
    this: dct:created "2019-05-03"^^xsd:dateTime ;
        pav:createdBy orcid:0000-0002-1144-6265 .
}
```

**Fig. 1.** Example of a typical nanopub [8]

In figure 1, the blue box illustrates the assertion graph, the smallest unit of a statement. The orange box represents the assertion provenance, and it describes how the assertion was generated and the methods used to compose it. The publication info graph comprises the information regarding the nanopub, like assertion subject, authors, rights information, and creation date, represented in the yellow box. Finally, the head graph represents the relation between the cited graphs and a triple to identify the file as a nanopub [8]. Kuhn et al. [15] report that the FDP and the LD approach are utterly necessary for scientific data management. Nevertheless, up to now, we do not yet have a robust, portable set of technologies to manage and steward data resources.

## 2.2 The impact of poor data curation in genomics

Genbank[1], Uniprot[2], ENA[3], and DDBJ[4] are traditional genomic databases used worldwide. Despite that, they have well-known issues reported by researchers [3]. For instance, GenBank is one of the most used. Despite being constituted by many databases, it still lacks several data controls and management issues [16].

Historically, data modelers or data experts did not contribute to designing genomic databases. Consequently, today's researchers who need to reuse genomic data (a.k.a, data reusers) often make intense efforts to discover and obtain datasets. Usually, they extract and test data trustworthiness by doing hand-made scripts. However, they frequently need to interoperate datasets with other databases and even refactor, parse or filter the data. Thus, duplication, inconsistency, inaccuracy, incompleteness, and outdatedness are frequent labour-intensive problems genomic researchers face [4].

Our previous work [17] has investigated data issues in some genomic databases, considering two data reusers personas: humans and machines. We scrutinised the FDP as the theoretical foundation to compose an evaluation framework for these genomic databases. The framework focused on the features of digital genomic objects to facilitate their discovery, access, interoperability, and reuse by data reusers. We created an evaluation framework on seven well-known genomic databases; the experimental results were below their beliefs and expectations. For instance, GenBank was only the sixth database compliant with the evaluation process, showing poor interoperability and reusability results for the machine-readable scenario [17].

## 2.3 Related Works

Several approaches addressing reusability for digital objects are present in the literature. One exemplary project is the nanopublication dataset of monogenic rare gene-disease associations (DisGeNET) [18]. It contemplates more than one million nanopubs referring to manually curated disease-gene associations following the DisGeNET model. The nanopubs are generated by automatically extracting the existing DisGeNET dataset, making them machine-automatable and ensuring immutable, permanent, and verifiable digital objects.

WikiPathWay is another nanopub project that stores assertions of biological pathway models, including metabolic, signalling, and genetic pathways [19]. The project's curators highlighted that nanopubs increase findability and reusability due to data provenance and the adoption of Globally Unique and Persistent Identifiers (GUPI).

---

[1] https://www.ncbi.nlm.nih.gov/genbank/
[2] https://www.uniprot.org/
[3] https://www.ebi.ac.uk/ena/browser/home
[4] https://www.ddbj.nig.ac.jp/index-e.html

NeXtProt employs nanopubs focusing on knowledge integration. In this case, human proteins (meta)data are extracted from the UniProtKB/Swiss-Prot database (one of the most used biological databases) and incorporated into the nanopub statements with other relevant information [20]. This approach focuses on provenance to increase discoverability and to establish quality thresholds. Provenance plays a crucial part in the judgment of data reliability, mainly when dealing with machine-actionable processes.

Prospective and retrospective data provenance has an essential role for nanopubs, representing two graphs (provenance and publication info graphs) of the four required in the original schema. Developing a nanopub schema that devotedly reflects the provenance associated with the data demands some effort when attending highly consolidated standards such as PROV-O.

Nevertheless, Asif et al. [11] state that a substantial part of nanopublications does not guarantee provenance and still has obstacles associated with semantic interoperability. The non-use of a suitable methodology, the possible limited data and expertise about the related domain, and the concepts of nanopubs may compromise the captured data and the nanopub itself. Additionally, the authors state that many nanopubs misunderstand the authoring roles, as they do not cite the assertion author(s), only the author(s) of the nanopub. In addition, the authors argue that a more profound analysis during the development and generation of data models for nanopubs is needed.

A well-founded reusability and interoperability approach for machine-readable processes may have a critical role in advancing knowledge discoverability, preventing information loss, reducing time, labour, and cost for data producers [21]. Initiatives like the FAIR principles are not new, although they can create a more robust and concise data reuse scenario for the existing data dilemmas. The convergence of semantic-related concepts and technologies is essential to mature this scenario and make data more understandable. The use of ontologies, controlled vocabularies, LD technologies, and other theories can provide this context [21, 22].

This paper proposes an approach to provide a well-established data model for GenBank genomic data. As previously mentioned, GenBank was one of the lowest in compliance with FAIR principles. Thus, we choose it as a use case. As pointed earlier, there is necessary to build a more reliable data environment for data interoperation and reuse.

Nanopubs can be very useful to circumvent the issues mentioned in this section. Nevertheless, there is still room for improvements in the data provenance and semantics of the nanopub data. Based on this, we propose an enhanced nanopub model to manage genomic data centered on semantic interoperability and provenance-aware concerns.

## 3    GAP (Genome Assembly nanoPublication)

The GAP approach combines related approaches to store GBAD data, emphasising data provenance, semantic interoperability, and machine-readable formats. We adopt the nanopublication approach to propose the GAP data model due to its naturalness of generating a more interoperable, reusable, and flexible data environment. To compose the GAP model, we needed to understand the meaning of each (meta)data stored and the data reusability scenario. As a first step, a genomic specialist contributed to better acknowledging the data and composing a data draft to generate the GAP model. Next, we recognised that supplementary metadata associated ought to be added to the schema.

These metadata stand for identifying the related genomic organism of GBAD (stored in NCBI taxonomy database) and articles that cite the data stored in GBAD (stored in PubMed articles database). This supplementary metadata is crucial to understand the stored information better and to increase knowledge discoverability.

To support that, we upgraded the GAP model to three different levels of nanopubs schemas (figure 2): (i) GBAD nanopub schema, referring to the metadata stored in GBAD (in green); (ii) the Organism nanopub schema referring to the organism taxonomy metadata (in purple); and (iii) the Article nanopub schema, related to scientific publications that cite the stored assembly data (in orange).

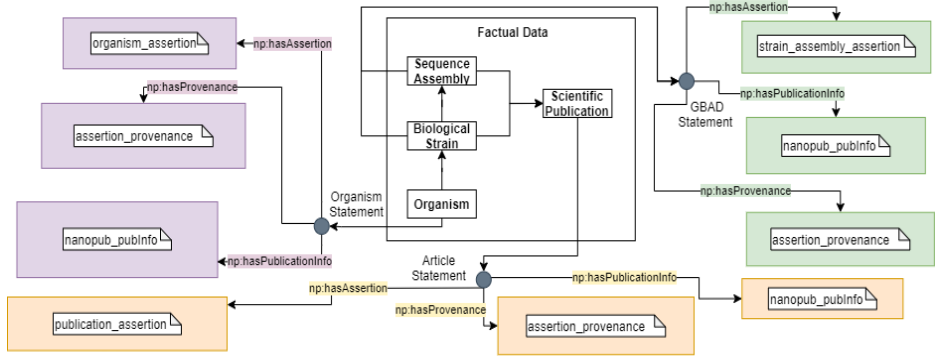

**Fig. 2.** Nanopubs schema diagrams

The next step was to specify how to represent the metadata. Besides, how this should be implemented adequately and which ontologies could be used to reproduce the semantical environment of GBAD data. To support enhanced semantics, the model used domain-controlled vocabularies. We adopted GUPI, avoided data literals (only used in particular cases such as publisher names), and provided a schema of nanopubs concentrating on their metadata provenance.

We used a total of 20 controlled vocabularies to deal with the complexity of representing three levels of the GAP model. The main ontologies to represent the biological data were the Semantic science Integrated Ontology (SIO) [23], the EMBRACE Data And Methods (EDAM) ontology [24], and the National Cancer Institute Thesaurus (NCIt) [25]. These ontologies are some of the most used when representing biological data. In addition, we adopted seven other biological ontologies to support the terms associated with GBAD metadata. Another significant point is the composition of the nanopubs schema for scientific articles. In this case, we utilised four additional ontologies related to semantic aggregation for the domain of scientific publications.

During the composition of the three nanopubs schemas, we established vocabularies to describe provenance and metadata in the nanopublication graphs. To compose and control provenance and publication info on nanopub graphs, we follow the W3C standard PROV ontology (PROV-O) [12]. PROV-O provides clear upper classes, relationships, and restrictions to frame any provenance. The nanopub graphs follow the Entity, Activity, and Agent concepts defined by the PROV model.

In addition, we used the Provenance, Authoring, and Versioning (PAV) ontology to describe authorship, curation, and digital creation of online resources [26], and we used DC terms for general metadata. Table 1 shows a summary of the most used semantic controlled vocabularies in the three-level data model. We use RDF to compose the nanopub data models to represent Genbank (meta)data. The three-level data model is serialised in TriG syntax to follow the pattern of the nanopub concept.

**Table 1.** Prefixes and descriptions of the most common namespaces

| Biological Controlled Vocabularies | | Provenance and Metadata Controlled Vocabularies | | Scientific Publication Controlled Vocabularies | |
|---|---|---|---|---|---|
| *Prefix* | *Namespace* | *Prefix* | *Namespace* | *Prefix* | *Namespace* |
| sio | Semanticscience Integrated Ontology | np | Nanopublication | prism | Publishing Requirements for Industry Standard Metadata |
| ncit | National Cancer Institute Thesaurus | rdfs | RDF Schema | cito | Citation Typing Ontology |
| edam | EMBRACE Data and Methods | xsd | XML Schema | fabio | FBBR-aligned Bibliographic Ontology |
| pato | The Phenotype and Trait Ontology | prov | Provenance Ontology | data | DataCite Ontology |
| so | Sequence Ontology | pav | Provenance, Authoring and Versioning | | |
| fbcv | FlyBase Controlled Vocabulary | dcterms | DCMI Metadata Terms | | |

Figure 2 illustrates that the different levels of the model are complementary. Associations were established within the nanopubs by referencing GUPI. Furthermore, it was essential to transform most data literals into identifiers to compose a semantically interoperable environment. Kuhn et al. [27] cite that the fewer literals there are in nanopublications, the better the use of machine-actionable processes will be.

Figure 3 illustrates the core section of the data model, the GBAD nanopub schema (following the colour schema present in figure 1). The complete versions of the Organism, GBAD, and Article nanopubs data models are available in a Github repository[5].

The GBAD nanopub assertion graph (blue box) stores all the (meta)data related to an organism assembly. All the (meta)data were transformed following the domain ontologies to obtain a fully semantic understandable scenario. In order to associate the related metadata present in the Organism nanopub, the SIO concept "SIO_000628" was used. This concept states a reference between two correlated nanopubs, the GBAD nanopub and the Organism nanopub (in this case, represented as "org_npub:5693").

The provenance graph (red box) comprises four different blocks of triples obeying concepts of PROV-O ontology, two entities generated by an automatic assertion and attributed to the submitter, the National Center for Biotechnology Information (NCBI), the owner of GenBank databases. To generate direct access to original and other related data, PROV concept "prov:hadPrimarySource" is used to connect to the FTP directory of the stored assembly data, which contains raw data, Genomic GenBank Format (.GBFF), and other associated files.

---

[5] https://github.com/MatheusFeijoo/Genome-Assembly-nanoPublication

The publication info graph (yellow box) follows the same PROV-O concepts and stores the nanopub provenance in five distinct triples blocks of Entity, Activity, and Agent. Besides data for nanopub provenance, this schema differentiates the authors of the assertion from the curators of the nanopublication (using ORCID unique digital identifier). Additionally, in "dcterms:dateSubmmitted", we prioritise using URIs instead of literals to identify related dates. By doing this, we may apply much faster filters when accessing the nanopubs by machine-readable processes [27].

**Fig. 3.** GBAD nanopub example in RDF/TriG notation using GAP data model.

Besides the assertion graph, the Organism and Article nanopub models follow the same head, provenance, and publication info template graphs present in the GBAD example. Additionally, in their publication info graphs, all nanopub levels within the GAP model use the "dcterms:subject" concept. The adoption of this concept aims to achieve better information discovery at the three levels of nanopubs and distinguish the content present between the levels.

## 4  Computational Experiments

We designed and executed a computational experiment composed of scripts to scrape and transform the GenBank (meta)data to nanopublications to evaluate the nanopublication models. Subsequently, we analyse the results with the real scenario of extracting (meta)data from GenBank and compare it with well-known nanopub datasets.

### 4.1  From Scraping Genomic Data to Generating Nanopublications

We developed Python scripts to automatically access, scrape, crawl and transform the collected GBAD data using the created nanopub schemas. We used the scripts to generate nanopubs related to assembly data of *Trypanosoma cruzi* and *Leishmania* diseases to test the schema. Additionally, we used the Scrapy framework to crawl and scrape the data from GBAD and related databases. Figure 4 illustrates the conceptual processes and the steps designed to generate the nanopubs.

The initial step receives as input the URL referring to an assembly of the diseases stored in GBAD and the ORCID of the data curators. The script analyses and searches for existing nanopubs referring to the instance in GBAD and its organisms. If there are no nanopubs, the following step extracts (meta)data referring to this organism in NCBI databases and scrapes (meta)data referring to the GBAD.

Simultaneously, a routine begins scraping GBAD instances (meta)data associated with the given organism. From there, all related GBAD registers are concentrated and stored in a JSON backup file. The subsequent step scrapes the PubMed database, aiming to collect (meta)data from scientific papers referencing the gathered instances.

After completing the previous steps, three JSON files are produced, each related to the levels of the nanopubs model. The following step begins converting the collected (meta)data to nanopubs based on the created model with those files.

The first level, transform the referenced organism (meta)data following the data model. Then, add the provenance related to the extraction and generation of the nanopub. In this transformation, the script creates a GUPI referring to the nanopub following the NCBI Taxonomy id of the organism.

As the second level, the code converts the data collected from the GBAD. The GBAD nanopub model includes all possible fields that can be stored. However, GBAD data are entered manually, and the entire model was not used. In this conversion, the GUPI of GBAD nanopublication follows the id present on GBAD. The GUPI referring to the organism nanopub is inserted into the nanopub, as mentioned in section 3.

Finally, at the last level, data referring to scientific papers that cite the collected GBAD instances are transformed to nanopubs. To relate these nanopubs with the GBAD nanopubs, the GUPI created for the GBAD nanopubs is used. The PubMed *id* is used on the creation of the GUPI of this nanopub.

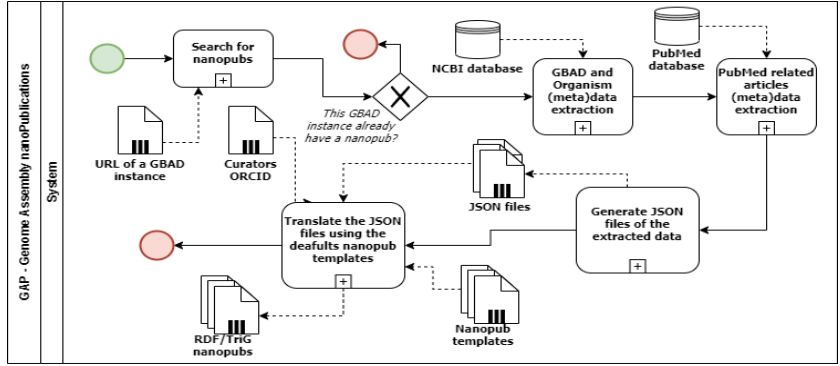

**Fig. 4.** The conceptual process to scrape, crawl and transform genomic data to nanopubs.

### 4.2 Nanopub Dataset

The nanopub dataset corresponds to two nanopubs of a given organism, 54 nanopubs of GBAD, and 14 nanopubs of scientific papers represented by 6.139 RDF triples. We can standardise the extracted (meta)data to be semantically understandable by machines by adopting the GAP data model. Only 12,99% of the (meta)data presented in the generated dataset are literals. In this case, they represent data about scientific publication tags, abstract text, names of provenance agents and insert data manually. We highlight that they are hard to format due to their heterogeneity.

It is possible to use SPARQL data query systems after converting the data to a machine-understandable format. Several works embrace extracting (meta)data from GenBank using unique approaches to perform the extractions [28, 29]. Each approach offers different results that are often not reused by other works.

This scenario can generate two problems: (i) increase in the amount of specific (meta)data and (ii) increase the effort to develop (meta)data extraction applications for specific GenBank scenarios. Our dataset is an example that circumvents these two problems, as it provides semantic-understandable (meta)data present in GenBank and eases the effort in producing new approaches for automatic (meta)data extraction. Additionally, the nanopub dataset can exploit the Linked Open Data (LOD) scenario to interoperate with other machine-readable data resources. LOD can be crucial in the genomic field by generating new knowledge, increasing information quality, and reducing gaps.

Another observation is about the provenance control improvement in nanopubs. Our results are comparable with the ones obtained by Asif et al. [11]. The authors evaluate the (meta)data quality of five well-known biologic nanopub repositories. Asif et al. state a lack of detailed data provenance in the provenance and publication info graphs of the analysed repositories, which may be related to the mechanism when capturing the data.

Additionally, Asif et al. found issues in the publication info graphs of the dataset. Data curators misunderstand when they designate the authors of a statement, curators, and creators of nanopublications [11]. When comparing our dataset with the types of problems presented by Asif et al. [11], none were detected. As the authors mention, it is necessary to use an adequate methodology to represent the extracted data completely. Our dataset achieves that by using the created GAP model that strictly follows the W3C PROV standard and is supported by a domain expert to precisely compose the data model that expresses the real semantic meaning of GenBank (meta)data.

## 5     Conclusions and Future Work

Data issues associated with semantics exist in many research areas; the genomic is just one example. Our investigation presented a novel data model approach to augment semantic interoperability for machine readability, indicating a feasible solution to these issues. We combined convergent concepts like LD, nanopublications, retrospective data provenance, FDP, and Web Ontology Language to mitigate machine data reuse issues. We developed a novel nanopub data model to transform and inject semantic meaning in one of the most used and problematic genomic databases, the GenBank.

Unlike the traditional scenario experienced by genomic data reusers, the GAP model increases the data semantic meaning and interoperability by using controlled vocabularies ontologies and consolidated models for data interchange. It can contribute to solving the data issues and constitutes a more robust way to represent data provenance in nanopubs. As future work, we intend to improve the data model to support: (i) better authoring, provenance representation, and interoperability between data silos; (ii) to scrape, transform and publish GenBank data in a machine-readable scenario; and (iii) to adapt the created model to use in other genomic databases; (iv) evaluate our approach in other knowledge domains.

## Acknowledgements

This study was financed in part by the National Council for Scientific and Technological Development (CNPq), Programa de Educação Tutorial (PET), Conselho Nacional de Pesquisa – Grant Number 315399/2018-0 and Coordenação de Aperfeiçoamento de Pessoal de Nível Superior – Brasil (CAPES) – Finance Code 001.